\documentstyle[11pt,newpasp,twoside,epsf]{article}
\def\edcomment#1{\iffalse\marginpar{\raggedright\sl#1\/}\else\relax\fi}
\reversemarginpar

\newcommand{\fig}[6]{
    \protect\centerline{
    \epsfxsize=#1\epsffile[#2 #3 #4 #5]{#6}
    }}

\begin{document}

\title{The Orientation of Accretion Disks Relative to Dust Disks in Radio Galaxies}
\author{Henrique R. Schmitt}
\affil{National Radio Astronomy Observatory, 1003 Lopezville, Socorro, NM\,87801, USA}

\setcounter{page}{1}
\index{Schmitt, H. R.}

\begin{abstract}
We study the orientation of accretion disks, traced by the position angle
of the jet, relative to the dust disk major axis in a sample of 20 nearby
Radio Galaxies. We find that the observed distribution of angles between
the jet and dust disk major axis is consistent with jets homogeneously
distributed over a polar cap of 77$^{\circ}$.
\end{abstract}

\section{Introduction}

Studying the orientation of accretion disks relative to host galaxy disks
can tell us about the inner structure of the galaxy and the regions
around the black hole. It is reasonable to assume that gas from the galaxy
disk is the source of fuel for the AGN, in which case we would expect the
accretion disk to be aligned with the host galaxy disk, and the jets to be
perpendicular to the host galaxy disk. The study of Seyfert galaxies
(Kinney et al. 2000) shows that their jets can have any orientation,
contradicting this scenario. In the case of Radio galaxies, which have elliptical
or S0 host, it is difficult to determine the orientation the host galaxy
disk. However, we solve this problem by using dust disks, a technique
similar to that used by Kotanyi \& Ekers (1979). We selected a sample of
20 Radio galaxies with dust disks and avoided sources with irregular dust
lanes. These results are presented in Schmitt et al. (2002), and similar
ones were found by Verdoes Kleijn et al. (this meeting).

\section{Results} 

The main results of this work are presented in Figure 1. The left panel of
this figure shows the distribution of angles $\delta$, the angle between
the jet and the dust disk major axis. It is clear that these radio galaxies
present a wide range of values, between $\sim15^{\circ}$ and 90$^{\circ}$.
This suggests that the jets are not aligned perpendicular to the dust disks.
However, the results from Figure 1 represent only a two-dimensional projection
on the plane of the sky, and we are interested on the intrinsic,
three-dimensional orientation of jets relative to the dust disks symmetry
axis. Using the techniques developed by Kinney et al. (2000), we have that
for a given pair of angles ($i$,$\delta$), where $i$ is the inclination of
the dust disk relative to the line of sight, it can be shown that the jet
will never be at an angle closer than $\beta_{min} = \sin ^{-1} (\sin i \cos
\delta$) relative to the disk symmetry axis. The right panel of Figure 1
presents the cumulative distribution of $\beta_{min}$'s for our sample,
where we can see that it shows a large distribution of values, indicating
that for half of the sample the jet and disk axis have to be misaligned by
more than 20$^{\circ}$, being larger than 53$^{\circ}$ for one of the
galaxies.

\begin{figure}[t]
    \fig{11cm}{20}{150}{585}{430}{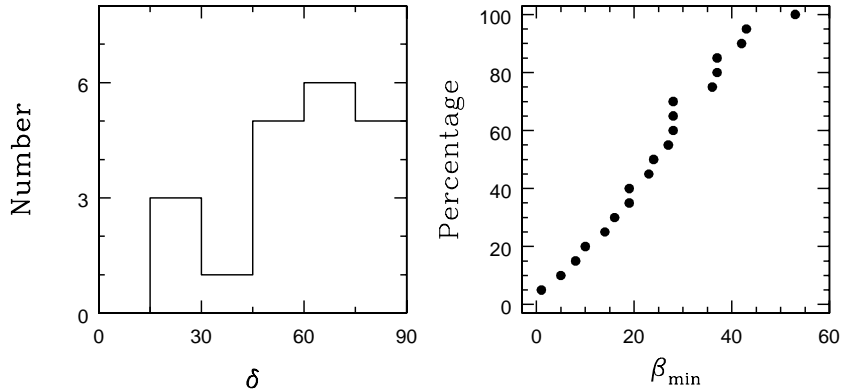}
\caption{Left: The distribution of angles $\delta$. Right: cumulative
distribution of angles $\beta_{min}$.}
\end{figure}

Besides studying the distribution of angles $\beta_{min}$, we also used the
statistical techniques developed by Kinney et al. (2000) to determine which
distribution of angles $\beta$ best represent the observed
distribution of $\delta$'s. We find that a homogeneous distribution of jets
over a polar cap of $77^{\circ}$ can represent the data at the 5\% level.
However, jets seem to avoid lying along the dust disks, which may be due
to the fact that they would not be able to propagate along that direction,
or would possibly even destroy these disks.

Kinney et al. (2002) and Schmitt et al. (2002) discuss several possibilities
to explain these results (see also Pringle, this meeting).
Misaligned inflow of gas towards the nucleus, from minor mergers,
or the accretion of individual molecular clouds do not seem to explain the
observations, since they would resulting in misaligned jets from the VLBI to the
VLA scales, which is not observed. Other possibilities would be the
warping of the accretion disk by self-irradiation instability (Pringle 1996),
by the Bardeen-Petterson effect, or by a misaligned gravitational potential
around the nucleus. These explanations also present some problems and require
further study.

\acknowledgements
I would like to thank my collaborators in this project: Jim Pringle,
Cathie Clarke and Anne Kinney. The National Radio Astronomy Observatory
is a facility of the National Science Foundation operated under
cooperative agreement by Associated Universities, Inc.

\end{document}